\documentclass[12pt]{article}
\usepackage{graphics}
\input epsf

\newcommand{\be}{\begin{eqnarray}}
\newcommand{\ee}{\end{eqnarray}}

\begin{document}
\title{Parity and Time Reversal in $J/\Psi$ Decay}
\author{L. Clavelli\footnote{lclavell@bama.ua.edu}, 
T. Gajdosik\footnote{garfield@bama.ua.edu}, 
and I. Perevalova\footnote{perev001@bama.ua.edu}\\
Department of Physics and Astronomy, University of Alabama,\\
Tuscaloosa AL 35487\\}
\maketitle

\begin{abstract}
     With the prospect of large numbers of $J/\Psi$ decay events becoming
available in the near future,  it is interesting to search for symmetry
violating effects as probes of new physics and tests of the
standard model.  $J/\Psi$ decay events could provide the first observation
of weak effects in otherwise strongly decaying particles.  We calculate
a T odd asymmetry in the $J/\Psi$ decay into photon plus lepton pair due to
Z boson exchange.  Extensions to hadronic final states are also discussed.
\end{abstract}
%\pacs{PACS numbers: 11.30.Pb, 12.60.J, 13.85.-t}
\renewcommand{\theequation}{\thesection.\arabic{equation}}
\renewcommand{\thesection}{\arabic{section}}
\section{\bf Introduction}
\setcounter{equation}{0}

In the US and China it is possible that current accelerators could be
adopted to become $J/\Psi$ factories with an initial annual yield of
$10^9 J/\Psi$.
Such machines could provide the first laboratory for the interplay of
weak and strong forces leading to parity and ultimately time reversal
violating effects in the hadronic final states.  Although parity odd
correlations should be easily observable in such machines, the observation
of true standard model time reversal violations will be quite challenging
and probably will require further luminosity upgrades beyond those of
the first generation machines. Some consideration of weak effects in
$J/\Psi$ decay have already been made in final states other than those
considered here \cite{weak}.
 Tests of T violation will require a careful
separation of pseudo-time-reversal violation coming, not from T violating
terms in the Lagrangian, but from final state interactions and particle
width effects.
\par
 In leptonic decays final state interactions are expected to be
negligible.
However the fact that unstable particles decrease in amplitude
rather than increase causes a possible T odd asymmetry to appear in leptonic
final states even in the absence of T odd terms in the Lagrangian.  In the
current article we calculate the $J/\Psi$ decay to photon plus lepton pair
including Z boson intermediate states.
\par
     In section II we calculate the standard model electroweak decay of the
$J/\Psi$ to photon plus lepton pair and present the distribution in photon
energy.  Section III is devoted to a discussion of the T odd asymmetry defined
by a forward-backward asymmetry of the normal to the plane defined by the
lepton pair.  In the current calculation this asymmetry is proportional to the
Z width and is below the sensitivity of the first generation $J/\Psi$ factory
proposals.  Nevertheless it serves as a background calculation for possible
new physics searches such as an electric dipole moment of the charm quark
as well as a model for larger though theoretically more
complicated hadronic final state effects.
     Conclusions and some discussion of related opportunities at $J/\Psi$
factories are presented in section IV.
\section{\bf
Standard Model amplitudes for $J/\Psi \rightarrow$ photon plus lepton pair.
}
\setcounter{equation}{0}
\par

\begin{figure}[tb]
\begin{center}
\begin{picture}(470,160)(0,0)
%\graphpaper[10](0,0)(480,130)
\put(-5, 5){\mbox{\resizebox{480pt}{!}{\includegraphics{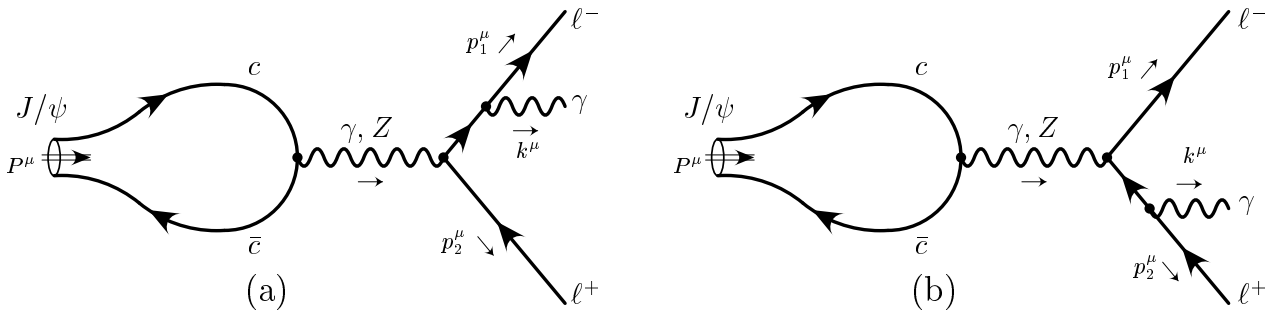}}}}
\end{picture}
\end{center}
\vspace*{-1cm}
\caption{Feynman diagrams with vector couplings only on the quark line}
\label{fig1}
\end{figure}

     The lowest order electroweak Feynman graphs for $J/\Psi$ are shown in
figures 1 and 2.  C invariance requires that in the amplitude ${\cal M}_2$
the lepton pair is produced by an intermediate $Z$ boson with axial coupling
at the charm quark line.  Therefore, in figure 2, the virtual photon and the
vector coupling of the $Z$ do not contribute.  The matrix element
is 
\be
 {\cal M}_2 = \frac{4 {\cal M}_0 e_q a_{Aq} i \varepsilon(\epsilon , \lambda
           ,\epsilon^\gamma,k)}{k \! \cdot \! P D_Z(q)} {\overline u}(p_1)
           \gamma_\lambda (a_{V l}+a_{A l} \gamma_5) v(p_2)
\ee
Here we neglect the lepton mass in numerator factors.
We have defined in terms of the totally antisymmetric Levi-Civita symbol
\be
    \varepsilon(a,b,c,d) = \varepsilon_{\mu\nu\alpha\beta} a^\mu b^\nu c^\alpha d^\beta
\enspace .
\ee
We employ the Bjorken-Drell sign conventions in which
\be
    \varepsilon_{0123}=+1
\enspace .
\ee
The electric charges of the quark and
lepton are $e_q =2/3, e_l = -1$.
The Z couplings to a fermion f are written
\be
{\cal L} = e Z_\lambda {\overline f} \gamma_\lambda (a_{Vf}+a_{Af}\gamma_5)f
\ee
with standard model values
\be
  a_{Vf}&=&(I_{3f}-2 e_f (\sin \theta_W)^2)/\sin (2\theta_W)\\
  a_{Af}&=&-I_{3f}/\sin(2 \theta_W)
\ee
We write for the denominator of the Z propagator
\be
     D_Z(q) = M_Z^2 - q^2 - i M_Z \Gamma_Z
\ee
Here $P, p_1, p_2, k$ respectively are the four
momenta of the $J/\Psi$,
the negatively charged lepton, the positively charged (anti) lepton,
and the photon.  In addition we have put $q = p_1+p_2$.
We use the non-relativistic color singlet model for the $J/\Psi$ where the
quark momentum and mass in the $J/\Psi$ are $P/2$ and $M/2$ respectively.
The model, based on the positronium formalism,
was extended to quarkonia some twenty years ago and has
been sucessfully employed in a number of applications~\cite{CSM}.
The non-relativistic quark model specifies the insertion of the wave function factor
\be
 F_{W} = \frac{\Psi(0)T^0}{\sqrt{12 M}}(\gamma \! \cdot \! P + M)  \gamma \! \cdot \! \epsilon
\ee
into the charm quark trace.  In Figure 1 a trace is taken over the Dirac and
color matrices, $T^0$ being the $3\times 3$ unit matrix in color space.

The wave function at the origin is then related to the electronic decay rate of the $J/\Psi$.
\be
   |\Psi(0)|^2 = \Gamma(J/\Psi \rightarrow e^+ e^- ) M^2 /(16 \pi \alpha^2
e_q^2) \approx .0044 GeV M^2
\ee
The wave function at the origin squared has dimensions $[E]^3$. The empirical
scaling with mass is indicated allowing our result to be also extended to
other vector quarkonia.
In terms of the fine structure constant, $\alpha$, and the $J/\Psi$ wave function at the
origin, the normalization constant ${\cal M}_0$ is given by a trace around the quark
line in Figure 1 together with a collection of coupling constants.  Namely
\be
  {\cal M}_0 \epsilon_\mu = 2 (4 \pi \alpha)^{3/2} Tr F_{W} \gamma_\mu
  = (4 \pi \alpha)^{3/2} {\sqrt {3M}} \Psi(0) \epsilon _\mu
\ee

\par
     The corresponding amplitude for the Feynman graphs of Fig. 1 is
\be
 {\cal M}_1 = {\cal M}_0 e_l \overline{u}(p_1) \left( 2 \left(
   \frac{p_2 \! \cdot \! \epsilon^\gamma}{p_2 \! \cdot \! k} -
 \frac{p_1 \! \cdot \! \epsilon^\gamma}{p_1 \! \cdot \! k} \right) \gamma \! \cdot \! \epsilon+
   \frac{\gamma \! \cdot \! \epsilon \, \gamma \! \cdot \! k \, \gamma \! \cdot \! \epsilon^\gamma}
        {p_2 \! \cdot \! k}-
   \frac{\gamma \! \cdot \! \epsilon^\gamma \, \gamma \! \cdot \! k \, \gamma \! \cdot \! \epsilon}
        {p_1 \! \cdot \! k} \right) \nonumber \\ 
    \cdot (A + B \gamma_5) v(p_2)
\ee
Here $A$ contains the vector couplings of the photon and Z while $B$ is the
axial term.
\be
\begin{array}{rl}
    A = &-\frac{e_q e_l}{M^2} + \frac{a_{Vq} a_{V l}}{D_Z(P)}\\
    B = &\frac{a_{Vq} a_{Al}}{D_Z(P)}
\end{array}
\ee
     In electron-positron annihilation the $J/\Psi$ is produced transversely
polarized.  Consequently the average over the $J/\Psi$ helicities is given by
\be
\nonumber
 \frac{1}{2} \sum _\lambda \epsilon_\beta (\lambda) \epsilon^\ast_\gamma (\lambda)& =&
  \frac{1}{2} \left(- g_{\beta \gamma} +\frac{P_\beta P_\gamma}{M^2}
   -\frac{\delta_\beta \delta_\gamma}{M^2}\right)\\
  & =& - \frac{1}{2} g_{\beta \gamma} + \frac{{p^e}_\beta {\overline{p^e}}_\gamma
         + {\overline{p^e}}_\beta {p^e}_\gamma }{M^2}
\label{spinsum}
\ee
where $P$ is the $J/\Psi$ momentum and
$p^e, \overline{p^e}$ are the initial state electron and positron momenta
respectively.
In addition we have put
\be
    \delta = p^e-p^{\overline e}
\enspace .
\ee
Because of the symmetry of the spin summation one can choose a basis in which
the $J/\Psi$ polarization vector is real.

     We neglect here a small departure from transverse polarization due to
production through the $Z$ whose effect on the distributions we calculate
is sub-leading in $(M/M_Z)^2$.  
Similarly we treat the $J/\Psi$
as a pure vector state neglecting any possible weak
mixing with the axial vector
quarkonium state that is several hundred $MeV$ higher in mass.
To adequately treat these $P$ wave states one must allow some relative
momentum between the quarks in the charmonium wave function.

The matrix elements squared summed over final state spins are then
\begin{eqnarray}
 |{\cal M}_1|^2 &=& \frac{16 (|A|^2+|B|^2)}{p_1\cdot k \, p_2 \! \cdot \! k}
    {\cal M}_0^2 e_l^2 \left( - \epsilon \! \cdot \! \epsilon
    \left[(p_1 \! \cdot \! P)^2 + (p_2 \! \cdot \! P)^2 \right]
    -M^2 \left[ (p_1 \! \cdot \! \epsilon)^2+(p_2 \! \cdot \! \epsilon)^2
    \right] \right)\\
 |{\cal M}_2|^2 &=& \frac{128 {\cal M}_0^2 e_{q}^2 a_{Aq}^2 (a_{V l}^2
   + a_{A l}^2)}{ (P \! \cdot \! k)^2 |D_Z(q)|^2} \left(
   - \epsilon \! \cdot \! \epsilon \, p_1 \! \cdot \! k \, p_2 \!
     \cdot \! k + k \! \cdot \! \epsilon
   \left[p_1 \! \cdot \! \epsilon \, p_2 \! \cdot \! k + p_2 \!
   \cdot \! \epsilon \, p_1 \! \cdot \! k
   \right] \right)
\ee
\be
2 Re{\cal M}_2{\cal M}_1^\ast =
 \frac{64 {\cal M}_0^2 a_{Aq} e_l e_q}
       {k \! \cdot \! P}
    &\times&
     \left(  D_2\varepsilon(P,\epsilon,p_1,p_2)
          \left( \frac{p_1 \! \cdot \! \epsilon}{p_1 \! \cdot \!
              k}+\frac{p_2 \! \cdot \! \epsilon}{p_2 \! \cdot \! k}
          \right)
     \right. \nonumber\\
 &  &  + \left. F_2
           \left( \epsilon \! \cdot \! \epsilon
                \left[ k \! \cdot \! P\,-M^2
                \right]
                   - k \! \cdot \! P
                \left[ \frac{(p_1 \! \cdot \! \epsilon)^2}{p_1 \! \cdot \! k} +
                  \frac{(p_2 \! \cdot \! \epsilon)^2}{p_2 \! \cdot \! k}
                \right] \right. \right. \nonumber\\
 &  &  \!\! + \left. \left. k \! \cdot \! \epsilon \left[ \frac{p_1 \! \cdot \! P
      p_1 \! \cdot \! \epsilon}{p_1 \! \cdot \! k} +
   \frac{p_2 \! \cdot \! P p_2 \! \cdot \! \epsilon}{p_2 \! \cdot \! k}
            \right]
         \right) \right)
\label{interferenceterm}
\ee
Here
\be
     D_2 =  Re \frac{i(a_{Vl}A^\ast + a_{Al} B^\ast)}{D_Z(q)}
\ee
and
\be
     F_2 = Re \frac {a_{Vl} B^\ast + a_{Al}A^\ast}{D_Z(q)}
\ee
\par

\begin{figure}[tb]
\begin{center}
\begin{picture}(470,200)(0,0)
%\graphpaper[10](0,0)(480,190)
\put(-5,10){\mbox{\resizebox{480pt}{!}{\includegraphics{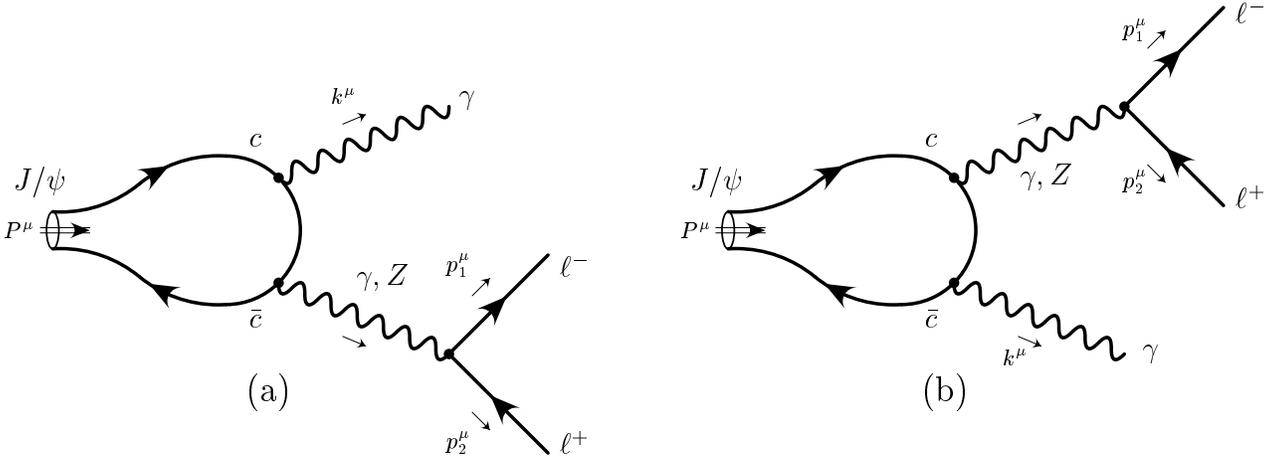}}}}
\end{picture}
\end{center}
\vspace*{-1cm}
\caption{Feynman diagrams with an axial coupling on the quark line}
\label{fig2}
\end{figure}

   Performing the spin average and partially integrating
over the phase space, we find the double distribution in photon energy and
angle relative to the incident electron as a sum of three terms corresponding
to the two direct terms and one interference term
\be
\nonumber
\frac{d^2\Gamma_{11}(\Psi \rightarrow l^+ l^- \gamma)}{dk_0 \,d(\cos \theta)}&=&
    E_{11} \left( 2 \sin^2{\theta_k}
              \left( \frac{M}{k_0} -2
              \right)
           \right.
    \\
&& \left. +(1 + \cos^2{\theta_k})
          \left( -1+ \ln{\frac{M(M-2 k_0)}{m_l^2}}
         \right)
         \left(-2+2\frac{k_0}{M}+\frac{M}{k_0}
         \right) \right)
\ee
where
\be
    E_{11} = 12 \Psi(0)^2 \alpha^3 (|A|^2+|B|^2) M
\enspace ,
\ee
\be
  \frac{d^2\Gamma_{22}(\Psi \rightarrow l^+ l^- \gamma)}{dk_0 \,d(\cos \theta)}=
  E_{22} \frac{k_0}{3 M}
    \left(1+ \sin^2{\theta}
      \left( \frac{1}{2}- 2\frac{k_0}{M}
      \right)
    \right)
\ee
where
\be
     E_{22} = 96 \alpha^3 M |\Psi(0)|^2 e_q^2 a_{Aq}^2  \frac{(a_{Vl}^2+a_{Al}^2)
             }{|D_Z(q)|^2}
\enspace ,
\ee
\be
%\nonumber
  \frac{d^2\Gamma_{12}(\Psi \rightarrow l^+ l^- \gamma)}{dk_0 \,d(\cos \theta)}=
   E_{12} \left( 1 + \left(\frac{1}{2}- 2 \frac{k_0}{M}
                     \right)\sin^2{\theta}
          \right)
\ee
with
\be
%\nonumber
   E_{12}= 2 a_{Aq} e_l e_q \frac{|{\cal M}_0|^2}{(2 \pi)^3} F_2
         = 48 \alpha^3 M  |\Psi(0)|^2 e_l e_q a_{Aq} F_2
\enspace .
\ee

     In spite of the presence of the Z, there is no forward backward
asymmetry in the photon direction. This is due to the assumed pure vector
nature of the $J/\Psi$ which results in its spin summation being
symmetric under the interchange $p^e \leftrightarrow p^{\overline e}$.

\section{\bf T odd Asymmetry}
\setcounter{equation}{0}
      In principle one could search for T violation in the $l^+ l^- \gamma$
final state in $J/\Psi$ decay.
A possible T-odd effect might be a forward backward asymmetry of the
normal to the plane of the final state leptons.
This could come from an electric dipole
moment of the charm quark or other new physics effects.  In the standard
model, there is no T violation in the Z couplings, however T-odd effects
could come from final state interactions or particle instabilities which
are not, of course, evidence for T violation in the Lagrangian.
By restricting our attention to non-hadronic final states, we avoid
strong final state interactions. However, since
unstable particles decrease in amplitude rather than increase, width effects
in propagators can mimic T violation.  Thus one could expect a T odd
asymmetry in $J/\Psi$ decay or in other processes due to intermediate
Z bosons.  Such effects will be proportional to the Z width to mass ratio.
In this section we calculate this T-odd asymmetry in the standard model
for its own sake and as a background to new physics searches.

    The appearance of the Levy-Civita tensor in \ref{interferenceterm}
is such a T-odd effect since the coeficient $D_2$ is proportional to the
Z width.

    Averaging this term over $J/\Psi$ polarizations using \ref{spinsum} we
have
\be
   |{\overline {\cal M}}|^2 = -\frac{-32 |{\cal M}_0|^2 a_{Al} e_l e_q D_2}{M^2}
       \frac{\varepsilon(P,\delta,p_1,p_2)}{k \! \cdot \! P}
       \left( \frac{p_1 \! \cdot \! \delta}{p_1 \! \cdot \! k} +
       \frac{p_2 \! \cdot \! \delta}{p_2 \! \cdot \! k} \right)
\label{intterm}
\ee
The contribution to the $J/\Psi$ decay rate is
\be
   d\Gamma = \frac{1}{2 M} |{\overline {\cal M}}|^2 d\Omega_3
\ee
with the standard Lorentz invariant phase space differential:
\be
  d\Omega_3 = \frac{d^3 k}{2 k_0} \frac{d^3 p_1}{2 p_{10}} \frac{d^3 p_2}{2 p_{20}} \frac{\delta^4(P-p_1-p_2-k)}{(2 \pi)^5}
\ee

One may use the $4D$ delta function to eliminate $p_2$ and afterwards it
is convenient to do the $p_1$ integral in the rest frame of $P-k$ with
the $k$ direction defining the $z$ axis and the $\delta$ direction taken to
be in the $xz$ plane.  Relative to this coordinate system the $p_1$
polar and azimuthal angles are $\theta_1$ and $\phi_1$ respectively.
The result can be put back into a manifestly Lorentz invariant form
allowing one to then do the $k$ integral in the $J/\Psi$ rest frame
with the $\delta$ (or $p^e$) direction defining the z axis.
In this latter frame the energy and polar angle of $k$ are $k_0$ and
$\theta_k$ respectively.  There can be no azimuthal dependence.

In terms of these variables and $w=2 k_0/M$ we have
\be
d\Omega_3 = \frac{M^2}{64 (2 \pi)^4} w dw d(cos(\theta_k)) d(cos(\theta_1)) d\phi_1
\ee
and
\be
   d\Gamma= \frac{M}{4} \frac{ |{\cal M}_0|^2 a_{Al} e_l e_q D_2}{(2 \pi)^4}
     \sqrt{1-w} \sin(\theta_k) \sin(\phi_1) dw d(cos(\theta_k)) d \theta_1
     d\phi_1 f
\ee
where
\be
     f = w \cos(\theta_k) \sin^2(\theta_1) - 2 \sqrt{1-w} \cos(\theta_1)
       \left( \cos(\theta_k) \cos(\theta_1) + \sin(\theta_k) \sin(\theta_1)
       \cos(\phi_1) \right)
\ee
A T-odd parameter is any function, y, of the final state momenta that is
odd in $\varepsilon(P,\delta,p_1,p_2)$.  In the $J/\Psi$ rest frame
this is proportional to the cosine of the angle between the normal to
the final state lepton plane and the initial state electron direction.
From an experimental point of view $\varepsilon(P,\delta,p_1,p_2)$ itself
is not the optimum variable to consider since
the corresponding distribution has an (integrable)
singularity at the origin.
A better variable from this point of view is
\be
      y = \sin(\theta_k) \sin(\phi) = - \frac{\varepsilon(P,\delta,p_1,p_2)}
      {M {\sqrt{2 p_1 \cdot k p_2 \cdot k p_1 \cdot p_2}}}
\ee
One may then define
\be
    \frac{d \Gamma}{dy} = \int d\Gamma \delta(y - \sin(\theta_k) \sin(\phi))
       \epsilon(k \cdot \delta)
\label{dgdy}
\ee
The factor $\epsilon(k \cdot \delta)$ implies that the events with
the photon momentum in the forward hemisphere relative to the
incident electron are to be counted negatively.  Without such a factor
the distribution would vanish identically upon integration over the
photon direction.

    The integral can be easily done by Monte-Carlo methods.  In figure 3
we plot $\frac{1}{\Gamma_{J/\Psi}}\frac{d \Gamma}{dy}$ versus $y$.
The distribution is too small to be seen at the first generation $J/\Psi$
factories where of order of $10^9$ events are projected.  This means that
\begin{figure}[tb]
\begin{center}
%%\begin{picture}(324,60)(0,0)
%\graphpaper[10](0,0)(480,60)
%%\put(  0,-5){\mbox{\resizebox{!}{4.5in}{\includegraphics{dgdyplot.eps}}}}
%%\end{picture}
%
\epsfxsize=4.5in %% 6.7in % actual
\leavevmode
\epsfbox{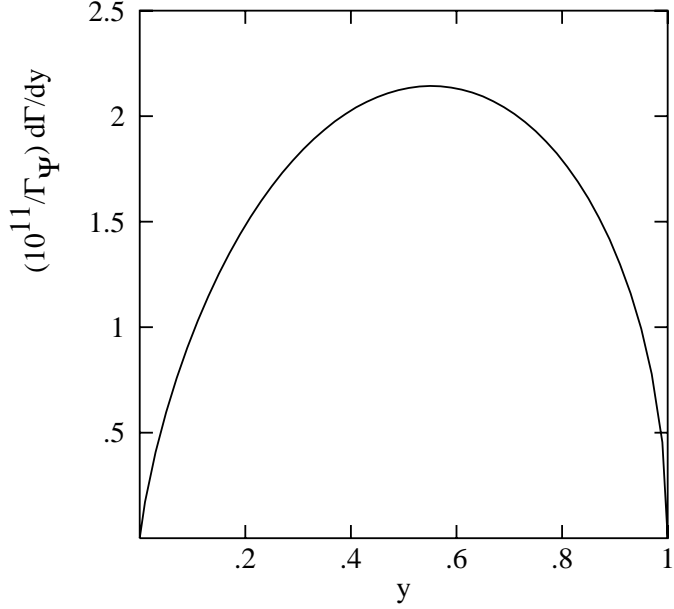}
\end{center}
\vspace*{-1cm}
\caption{ Asymmetry distribution (times $10^{11}$) in the region of
positive y (see text)}
\label{fig3}
\end{figure}
a search for a y asymmetry is sensitive to new physics such as, perhaps,
an electric dipole moment of the charm quark.  Most models for electric
dipole moments, such as supersymmetry, predict an effect proportional to
the mass of the fermion.  Thus the electric dipole moment of the charm
quark can be expected to be orders of magnitude greater than any effect
in the light quarks \cite{CGM}.  We leave a full calculation of the
T-odd asymmetry in $J/\Psi$ decay in the presence of a charm quark
electric dipole moment to a later paper.  At present
we know of no experimental limits on charm quark electric dipole moments.

It is clear from Eq.~(\ref{dgdy}) that the distribution is antisymmetric in y.
To leading order in $(1/M_Z)$ it can be written
analytically.
\be
    \frac{d \Gamma}{dy} =
    \frac{ 11 \alpha^3 |\Psi(0)|^2 a_{Vl} a_{Aq} (e_l e_q)^2 \Gamma_Z}{5 M_Z^3}
      y \ln \frac{1+{\sqrt{1-y^2}}}
      {1 - {\sqrt{1-y^2}}}
\ee
Integrated over positive  values of $y$ and
expressed relative to the $J/\Psi$ total width this is
\be
   \frac{1}{\Gamma_\Psi} \int_0^1 dy \frac{d\Gamma}{dy} =
  \frac{ 11 \alpha^3 |\Psi(0)|^2 a_{Vl} a_{Aq} (e_l e_q)^2 \Gamma_Z}{5 M_Z^3
  \Gamma_\Psi} = 1.617 \cdot 10^{-11}
\ee

\section{\bf Conclusions}
\setcounter{equation}{0}

     We have calculated the photon spectrum and angular distribution
in the standard model decay of the $J/\Psi$ into photon plus
lepton pair including the effects of an intermediate Z boson.
In addition we have plotted the T-odd asymmetry distribution of
the normal to the final state lepton pair relative to the beam
direction.  A small but non-zero effect is found proportional to
the $Z$
boson width.  The effect is proportional to the vector coupling
of the $Z$ to leptons which is suppressed by
the closeness of
$\sin^2(\theta_W)$ to $1/4$.  This suggests that a significantly
larger result might be found in photon plus hadronic jets.
Some discussion of such effects in exclusive channels. 
has already appeared \cite{Ma}.
In the photon plus leptons channel, 
the current calculation provides a good bound on standard model
backgrounds
to a search for genuine T violation in $J/\Psi$ decay due,
for example, to a
possible electric dipole moment of the charm quark for which no
experimental bounds have as yet been published.

{\em acknowledgments}

     This work was supported in part by the US Department of Energy
under grant number DE-FG02-96ER-40967.  LC thanks the physics
department of New York University for hospitality during part of
the period of this research.  Similarly, TG acknowledges the
hospitality of the Institute of Theoretical Physics and Astronomy in
Vilnius, Lithuania during the summer of 2001.  In addition
the authors acknowledge frequent discussions with Philip Coulter.

%%%%%%%%%%%%%%%%%%%%%%%%%%%%%%%%%%%%%%%%%%%%%%%%%%%%%%%%%%%

%%%%%%%%%%%%%%%%%%%%%%%%%%%%%%%%%%%%%%%%%%%%%%%%%%%%%%%%%%%


\begin{thebibliography}{ABCDEF}
\nonumber
\bibitem{weak} X.-G. He, J.-P.Ma, and B. McKellar, Phys.~Rev.~D{\bf 49},
4548 (1994)\\
\nonumber
A. Datta, P.J. O'Donnell, S. Pakvasa, and X. Zhang, Phys.~Rev.~D{\bf 60},
014011 (1999) \\
S. Nussinov, R.D. Peccei, and X. Zhang, hep-ph/0004153\\
X. Zhang, hep-ph/0010105\\
\nonumber
\bibitem{CSM} J.~H.~K\"uhn, J.~Kaplan, and E.~Safiani, Nucl.~Phys.~
{\bf B157}, 125 (1979)\\
\nonumber
W.~Y.~Keung, in Proceedings of the Cornell $Z^0$ Workshop, 1981
(unpublished), Phys.~Rev.~D{\bf 23}, 2072 (1981)\\
\nonumber
E.L. Berger and D. Jones, Phys.~Rev.~ D{\bf 23}, 1521 (1981)\\
\nonumber
L.~Clavelli, Phys.~Rev.~ D{\bf 26}, 1610 (1982)
\bibitem{CGM} L.~ Clavelli, T. ~Gajdosik, and W.~Majerotto,
Phys.~Lett.~ B{\bf 512}, 115 (2001)
\nonumber
\bibitem{Ma} J.-P. Ma, Nucl~Phys.~B{\bf 602}, 572 (2001)\\
\nonumber
J.-P. Ma and J.S. Xu, Phys. Lett.~B{510}, 161 (2001)
\end{thebibliography}
\end{document}